\documentclass[conference]{IEEEtran}
\usepackage{cite}
\usepackage{array}
\usepackage{epsfig}
\usepackage{amssymb}
\usepackage{amsfonts}
\usepackage{graphicx}
\usepackage{subfigure}
\usepackage[cmex10]{amsmath}
\newcommand{\vtheta}{{\boldsymbol \theta}}

\begin{document}
%
\title{Cooperative Radar and Communications Signaling: \\
The Estimation and Information Theory Odd Couple }

\author{
\IEEEauthorblockN{Daniel W.~Bliss}
\IEEEauthorblockA{
School of Electrical, Computer and Energy Engineering\\
Arizona State University\\
Tempe, Arizona USA}
}

\maketitle

\begin{abstract}
We investigate
cooperative radar and communications signaling.
While each system
typically considers the other system a source of interference,
by considering the radar and communications
operations to be a single joint system,
the performance of both systems can, 
under certain conditions,
be improved 
by the existence of the other.
As an initial demonstration,
we focus on the radar as relay scenario
and present an approach
denoted multiuser detection radar (MUDR).
A novel joint estimation and information theoretic
bound formulation is constructed
for a receiver
that observes communications and radar return 
in the same frequency allocation.
The joint performance bound is presented 
in terms of the communication rate
and the estimation rate of the system.
\end{abstract}

\IEEEpeerreviewmaketitle
\long\def\symbolfootnote[#1]#2{\begingroup\def\thefootnote{\fnsymbol{footnote}}
\footnote[#1]{#2}\endgroup}

\symbolfootnote[0]{
This work was sponsored in part by DARPA under the SSPARC program.
The views expressed are those of the author and do not
reflect the official policy or position of the Department of Defense 
or the U.S. Government.}

\section{Introduction}

Given the reality of the ever increasing strain 
on limited spectral resources,
radar and communications systems 
are in some cases being forced into an uneasy coexistence.
The typical assumption is that the existence 
of one type of system (either a radar or a communications system)
will degrade the performance of the other system.
Consequently, the systems are 
usually isolated temporally, spectrally, or spatially
in most operations.

\subsection{Background}

During the last decade, cognitive radio technologies 
\cite{Mitola06,BlissAWC2013}
have been considered that implement opportunistic spectrum 
sharing as they are able to sense under-utilized spectrum 
and adaptively allocate it to other users \cite{Yucek09}.
A similar coexistence problem is currently faced by radars 
as their performance deteriorates 
due to coexisting wireless communications systems.
Cognitive radars indicate initial
attempts to adapt intelligently to complicated environments
\cite{Guerci2010cognitive}.

Current research on the spectral
coexistence of radar and communications systems has mainly
involved concepts similar to cooperative sensing
\cite{Wang08,Nij11,Bhat12,Sar12,Sar12_b}.
Other methodologies that have been applied to the
radar-communications coexistence problem include
signal sharing \cite{Li12,Jam08} and waveform 
shaping \cite{Chen11,Yan11,Sod12,Surender10}.
In other research and applied systems
radars based on communication system waveforms
have been considered.
As an example,
operating the radar passively or parasitically
by using a broadcast communication system
has been investigated
(for example in \cite{Griffiths2008bistatic} and references therein).
Also,
radios that communicate with radar systems
by modulating the radar waveform
have been considered \cite{Bidigare2002Shannon}.

\subsection{Contributions}

The principal contribution
of the paper
is that 
we develop 
a novel performance bound formulation
to provide insight 
into the limits 
of coexisting radar and 
communications systems. 
For joint decoding and radar channel estimation
(which we denote multiuser detection radar: MUDR),
we allow the radar to demodulate and decode
the communications signal jointly with estimating its 
radar channel. 
Rather than have radar and communications
system performance degraded, the performance of both 
systems is potentially enhanced by the systems' interactions. 
In its most general form, the coexisting 
radar and communications system becomes a large heterogenous 
multistatic radars or 
statistical multiple-input multiple-output (MIMO) radar 
\cite{Chernyak1998fundamentals,Fishler04,ForsytheBook08} and 
simultaneously a heterogenous communication network
\cite{BlissAWC2013}.
These jointly cooperative systems
are only possible under certain theoretical constraints
that we begin to explore in this paper.

In this paper,
as a preliminary exploration,
we consider the limited scenario 
of a joint radar and communications relay.
In this case,
the node traditionally denoted ``radar" 
is also a communications relay
that jointly estimates the radar return
and receives a communication signal.
The radar waveform is then assumed 
to be a communications waveform.
Because of the advantages of the radar power, 
the performance of the communications 
between two or more nodes
is typically improved
by using the radar as a relay
compared to direct ground-to-ground communications.
The principal constraint
in performance of this system
is in simultaneous reception of the radar return
and communications signal,
and is therefore the main thrust of this work.



\section{Joint Estimation/Communications Bounds} 

In general,
much like network communications \cite{BlissAWC2013},
exact bounds are challenging.
However,
in certain cases,
such as the multiuser base station,
bounds are tenable.
We develop a generalization of the multiple-access receiver discussion
for the joint radar channel estimation and communications
reception.

\subsection{Multiple-Access Communications Analogy}
\begin{figure}[htb]
\centerline{
   \includegraphics[trim =20mm 17mm 2mm 15mm, clip,width=3.7in]{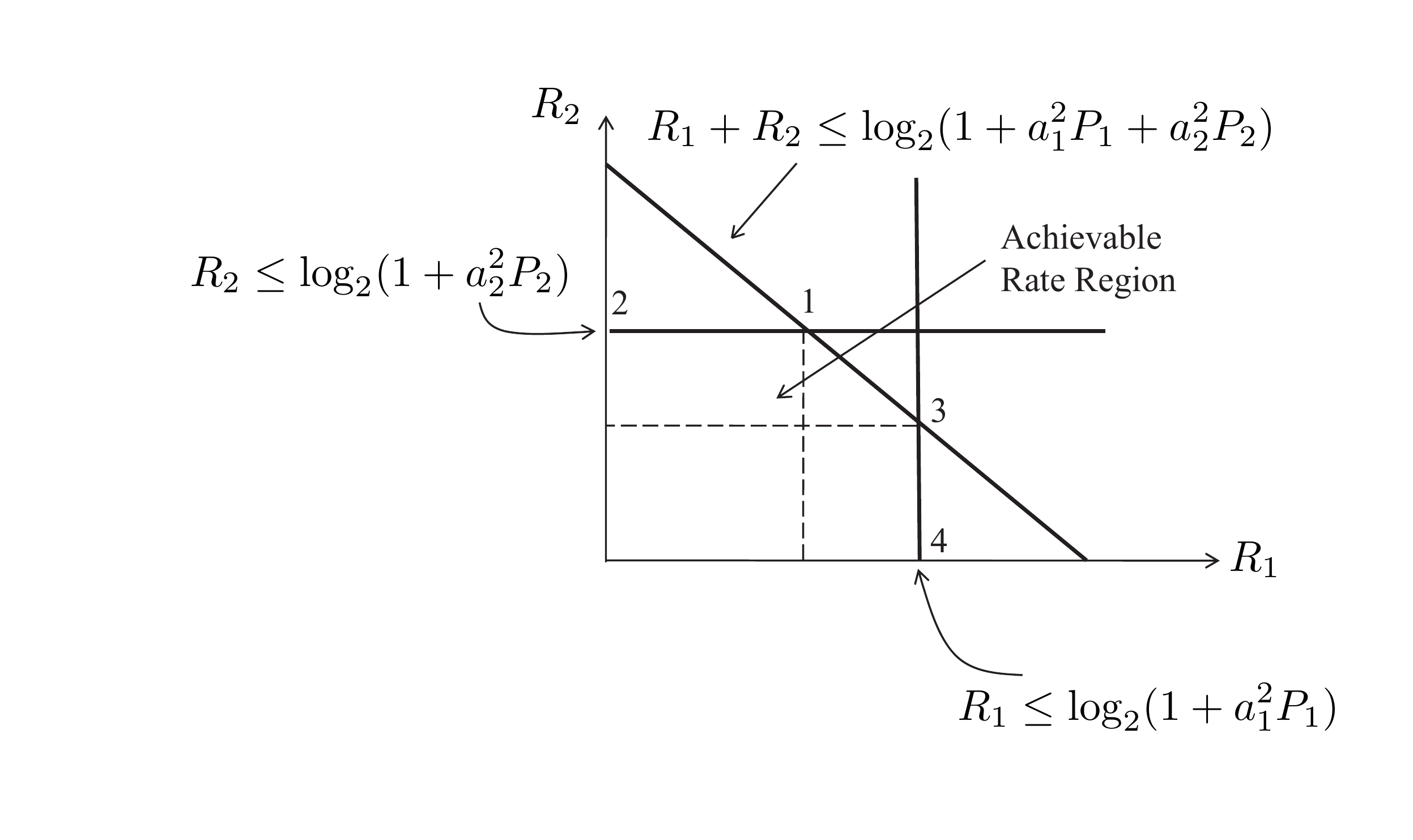}}
  \caption{Pentagon that contains communications multiple-access 
  achievable rate region.}
  \label{fig:pentagonOuterBound}
\end{figure}

For reference,
we review the multiple-access communications system 
performance bound \cite{Cover06,BlissAWC2013}.
In the multiple-access channel
that we discuss here,
we assume that two independent transmitters
are communicating with a single receiver.
The channel-attenuation-power product
for the two transmitters are given by
$ a_1^2 P_1$ and $a_2^2 P_2$,
respectively.
Their corresponding rates are denoted $R_1$ and $R_2$.
Assuming that power is normalized
so that the noise variance is unity,
the fundamental limits on rate are given by%
\footnote{Note: We assume complex baseband signals,
so there are two degrees of freedom;
thus, there is no ``1/2" before the $\log$ term}
\begin{align}
  R_1 &\le \log_2(1 + a_1^2 P_1) 
  \nonumber \\
  R_2 &\le \log_2(1 + a_2^2 P_2) 
  \nonumber \\
  R_1 + R_2 & \le \log_2(1 + a_1^2 P_1 + a_2^2 P_2) 
\end{align}
Vertices are found by jointly solving two bounds,
\begin{align}
  R_2 &= \log_2(1 + a_2^2 P_2) 
  \nonumber \\
  R_1 + R_2 - R_2 &= \log_2(1 + a_1^2 P_1 +  a_2^2 P_2) - \log_2(1 + a_2^2 P_2)  
  \nonumber \\
  R_1 &= \log_2\!\left(\frac{1 + a_1^2 P_1 +  a_2^2 P_2}{1 + a_2^2 P_2}\right)
  \nonumber \\
  \{R_1,\, R_2\}&=  \left\{ \log_2\!\left(1 + \frac{a_1^2 P_1}{1 + a_2^2 P_2}\right), \,
  \log_2(1 + a_2^2 P_2) \right\}
  \, ,
\end{align}
and
\begin{align}
  \{R_1,\, R_2\}&=  \left\{  \log_2(1 + a_1^2 P_1) ,
  \log_2\!\left(1 + \frac{a_2^2 P_2}{1 + a_1^2 P_1}\right) \,
 \right\}
  \, .
\end{align}
The region that satisfies these theoretical bounds
is depicted in Figure \ref{fig:pentagonOuterBound}.

Unfortunately,
this discussion serves only as a motivation 
because radar returns do not satisfy
the fundamental communications assumption
that they are drawn from a countable dictionary.
Consequently,
we do not expect that this form is directly applicable.
However,
by using a formalism similar to the communications
multiple-access bound,
we can gain insight into the simultaneous 
channel use by communications and radar.

\subsection{Joint Radar-Communications Notation}

Because there is a significant quantity of notation
in discussing this topic,
in Table \ref{tab:notation}
we present an overview of the important notation employed.

\begin{table}[htdp]
\caption{Survey of Notation.}
\begin{center}
\begin{tabular}{|c|p{6.5cm}|}
\hline
Variable & Description \\
\hline \hline
$\left< \cdot \right> $ & Expectation \\
$\| \cdot \|$ & L2-norm or absolute value \\
$B$ & Total system bandwidth \\
$z(t)$ & Observed signal including radar and communications \\
$\tilde{z}(t)$ & Observed signal with predicted radar return removed \\
$z_{\rm radar}(t)$ & Observed radar return \\
$s_{\rm radar}(t)$ & Unit-variance transmitted radar signal \\
$P_{\rm radar}$ & Radar power \\
$\tau_m$ & Time delay to $m^{th}$ target \\
$\tau_m^{(k)}$ & $k^{th}$ observation of delay for $m^{th}$ target \\
$\tau_{m, \rm pre}$ & Predicted time delay to $m^{th}$ target \\
$a_m$ & Combined antenna gain, cross-section, and propagation \\
$N$ & Number of targets \\
$T$ & Radar pulse duration \\
$N$ & Number of targets \\
$T_{\rm pri}$ & Pulse repetition interval \\
$\delta$ & Radar duty factor \\
$s_{\rm com}(t)$ & Unit-variance transmitted communication signal \\
$P_{\rm com}$ & Total communications power \\
$b$ & Communications propagation loss \\
$n(t)$ & Receiver thermal noise \\
$\sigma^2_{\rm noise}$ & Thermal noise power \\
$k_B$ & Boltzmann constant \\
$T_{\rm temp}$  & Temperature\\
$n_{\rm int+n}$ & Interference plus noise for communications receiver \\
$\vtheta$ & Set of nonspecific system and target parameters \\
$B_{\rm rms}$ & Root-mean-squared radar bandwidth \\
$\gamma$ & Radar spectral shape parameter \\
$B_{\rm com}$ & Communications-only subband \\
$B_{\rm mix}$ & Mixed radar and communications subband \\
$\alpha$ & Fraction of bandwidth for communications only \\
$\beta$ & Power fraction used by communications-only subband \\
$\mu_{\rm com}$ & Channel of communications-only subband \\
$\mu_{\rm mix}$ & Channel of mixed use subband \\
\hline
\end{tabular}
\end{center}
\label{tab:notation}
\vspace{-5mm}
\end{table}%

\subsection{Joint Radar-Communications Channel Model} 

In this section,
we consider bounds for the multiple-access communications
and radar return channel.
We employ a number of simplifying assumptions
for the sake of exposition;
however, generalizations are possible.
As an example,
we estimate the range, 
but assume that the target cross-section is known.
We assume that the targets are well separated
and the that return is modeled well by a Gaussian distribution
before pulse compression.
We assume that the range of any given
target is predictable
up to some Gaussian random process variation
(not be confused with estimation error).
We consider only the portion of time
during which the radar return overlaps
with the communications signal.
We assume that temporal uncertainty
of the random target process is within
one over the bandwidth.

For $N$ targets,
the observed radar return $z_{\rm radar}(t)$ as a function of $t$ is given by
\begin{align}
  z_{\rm radar}(t) &= 
  \sum_{m=1}^N a_m  \, s_{\rm radar}(t-\tau_m) \, \sqrt{P_{\rm radar}} + n(t) \, .
\end{align}
The zero-mean noise
is drawn from the complex Gaussian
with variance $ \sigma_{\rm noise}^2$,
\begin{align}
 \sigma_{\rm noise}^2 &= k_B \, T_{\rm temp} \, B \, ,
\end{align}
where $k_B$ is the Boltzmann constant,
$T_{\rm temp}$ is the absolute temperature,
and $B$ is the full bandwidth.
Range $r$ and delay $\tau$ are related by
\begin{align}
  \tau &= \frac{2 \, r}{c} \, ,
\end{align}
where $c$ is the speed of light.
The typical radar estimator 
attempts to estimate both the range
and the amplitude.
For the sake of discussion,
we focus on range estimation.
Similar developments 
can be found for amplitude estimation.
A reasonable estimator
(particularly if targets are well separated)
under the assumption 
that Doppler shifts are unresolvable
is given by
\begin{align}
  \hat{\tau}_m &= {\rm argmax}_{\tau_m} \int dt \, z(t) \, s_{\rm radar}^*(t-\tau_m)
  \, .
\end{align}
We assume that we are tracking the target,
and we assume the optimistic model
that we have some well understood expected value of the radar return
(based upon prior observations);
however,
there is some range fluctuation in the return
due to some underlying target process,
so that the next observation
is known up to some random Gaussian process variation $n_{\tau, \rm proc}$,
\begin{align}
  \tau_m^{(k)} &= \tau_{m, \rm pre}^{(k)} + n_{\tau, \rm proc} \\
  \tau_{m, \rm pre}^{(k)}&= f(k;T_{\rm pri},\vtheta)
  \, .
  \nonumber
\end{align}
The function $f(k;T_{\rm pri},\vtheta)$
is a prediction function
with parameters $T_{\rm pri}$,
which is the time between updates
(pulse repetition interval),
and $\vtheta$
which contains other parameters.
The variance of the process is given by
\begin{align}
   \sigma^2_{\tau, \rm proc}
  &= \left<\left\|\tau_m^{(k)} - f(k;T_{\rm pri},\vtheta) 
  \right\|^2\right>
  \, .
\end{align}
The observed signal at the receiver $z(t)$
at time $t$
in the presence of a communications signal
and the radar return is given by
\begin{align}
  z(t) &= \sqrt{P_{\rm com}} \, b \, s_{\rm com}(t)
   \\
  & \qquad + \sqrt{P_{\rm radar}} \, 
  \sum_{m=1}^N a_m  \, s_{\rm radar}(t-\tau_m) \, + n(t)  
  \nonumber
\end{align}
%

\subsection{Radar-Prediction-Suppressed Observed Signal}

For the sake of the communications system,
we can try to mitigate
unnecessary interference
by subtracting the predicted
radar return at the receiver\footnote{Note: 
this process would theoretically remove all clutter.}
\begin{align}
  \tilde{z}(t) &= \sqrt{P_{\rm com}} \, b \, s_{\rm com}(t) + n(t)
   \\
  & \qquad + \sqrt{P_{\rm radar}} 
  \sum_{m=1}^N a_m  [s_{\rm radar}(t\!-\!\tau_m)\!-\!s_{\rm radar}(t\!-\!\tau_{m, \rm pre})] 
  \, ,
  \nonumber
\end{align}
where here we dropped the explicit indication of pulse index $(k)$.
For small delay process variation,
we can replace the difference between
the waveforms at the correct and predicted delay 
with a derivative,
\begin{align}
  &s_{\rm radar}(t-\tau_m)-s_{\rm radar}(t-\tau_{m, \rm pre})
  \nonumber \\
  & \qquad = s_{\rm radar}(t-\tau_m)-s_{\rm radar}(t-\tau_m + n_{\tau, \rm proc})
  \nonumber \\
  & \qquad \approx 
  \frac{\partial s_{\rm radar}(t-\tau_m)} {\partial t}  
  \, n_{\tau, \rm proc} \, .
\end{align}
The observed signal
is then given by
\begin{align}
  \tilde{z}(t) &\approx \sqrt{P_{\rm com}} \, b \, s_{\rm com}(t) + n(t) 
  \nonumber \\
  & \qquad + \sqrt{P_{\rm radar}} \, 
  \sum_{m=1}^N a_m  \, \frac{\partial s_{\rm radar}(t-\tau_m)} {\partial t}  
  \, n_{\tau, \rm proc}  
  \, .
\end{align}
From the communications
receiver's perspective,
the interference plus noise
is given by
\begin{align}
  n_{\rm int+n}
  &\approx  \sqrt{P_{\rm radar}} \, 
  \left(\sum_{m=1}^N a_m  \, \frac{\partial s_{\rm radar}(t-\tau_m)} {\partial t}  
  \, n_{\tau, \rm proc} \right)\, + n(t)  
  \nonumber \\
  \sigma^2_{\rm int+n}
  &= \left<\| n_{\rm int+n}\|^2 \right> 
  \nonumber \\
  &= P_{\rm radar}\left( \sum_{m=1}^N a^2_m \, 
  (2 \pi)^2 \, B^2_{\rm rms} \, \sigma^2_{\rm proc} \right)
  + \sigma^2_{\rm noise}
  \, ,
  \label{eq:intPlusNoise}
\end{align}
where 
$B_{\rm rms}$ is extracted
by employing Parseval's theorem \cite{BlissAWC2013}.
The value $\gamma$ is the scaling constant
between $B$ and $B_{\rm rms}$ times $2 \pi$
that is dependent upon the shape of the radar waveform's power spectral density.
For a flat spectral shape, 
$\gamma^2 = (2 \pi)^2/12$.

\subsection{Radar Estimation Information Rate} 

An essential tool of this paper
is to consider the estimation information rate
(estimating delay in this case).
We develop this information rate by considering
the entropy of a random parameter being estimated
and the entropy of the estimation uncertainty of that parameter.
As an observation,
if the targets are well separated,
then each target estimation can be considered
an independent information channel.

\subsubsection{Estimation Entropy} 

To find the estimation entropy,
we find the delay estimation uncertainty 
for each target.
For circularly symmetric Gaussian noise,
we employ the complex Slepian-Bangs formulation 
of the Cramer-Rao bound \cite{Kay93,BlissAWC2013}.
The variance of delay estimation for the $m^{th}$ target 
(ignoring inter-target interference) 
is given by
\begin{align}
  \sigma^2_{\tau;\rm est} &= {\rm Var}\{\hat{\tau}_{m} \} 
  = \frac{1}{(2\pi)^2 \, B^2_{\rm rms} \, {\rm ISNR}} 
  \nonumber \\
  &= \frac{\sigma_{\rm noise}^2}
  {(2\pi)^2 \, B^2_{\rm rms} \, TB \, a^2_m \, P_{\rm radar}} 
  \nonumber \\
  &= \frac{k_B \, T_{\rm temp}}{\gamma^2 \, B \, (TB) \, a^2_m \, P_{\rm radar}} 
  \, ,
\end{align}
where ${\rm ISNR} = TB \, a^2_m \, P_{\rm radar}/\sigma_{\rm noise}^2$ 
indicates the integrated SNR,
and the thermal noise is given by
\begin{align}
 \sigma_{\rm noise}^2 &= k_B \, T_{\rm temp} \, B 
 \, .
\end{align}
Under the assumption of Gaussian estimation error,
the resulting entropy 
of the error is given by
\begin{align}
  h_{\tau, \rm est} &= \log_2[\pi \, e \,  \sigma_{\tau, \rm est}^2] 
  \nonumber \\
  &= \log_2\!\left[\pi \, e \, \frac{k_B \, T_{\rm temp} }
  {\gamma^2 \, B \, (TB) \, a_m^2 \, P_{\rm radar}} \right]
  \, .
\end{align}
%

\subsubsection{Radar Random Process Entropy} 

The entropy of the process uncertainty plus estimation uncertainty
under a Gaussian assumption for both
is given by \cite{Cover06,BlissAWC2013}
\begin{align}
  h_{\tau, \rm rr} 
  &=  \log_2\!\left[\pi \, e \, (\sigma_{\tau, \rm proc}^2 + \sigma_{\tau, \rm est}^2)\right]
  \, .
\end{align}
%

\subsubsection{Estimation Information Rate}

Consequently,
the mutual information rate
in terms of bits per pulse repetition interval $T_{\rm pri}$,
which is related to the integration period $T$ 
by the duty factor $T = \delta \, T_{\rm pri}$,
is approximately bounded by
\begin{align}
  R_{\rm est} &\leq \sum_m \frac{h_{\tau, \rm rr}  - h_{\tau, \rm est} }{T_{\rm pri}}
  = 
  \sum_m \frac{\delta}{T} \log_2\!\left(1+ 
  \frac{ \sigma_{\tau, \rm proc}^2}{\sigma_{\tau, \rm est}^2}\right) 
  \nonumber \\
  &= \sum_m B \log_2\!\left( 1 + 
  \frac{ \sigma_{\tau, \rm proc}^2 \, \gamma^2 \, B \, (TB) \, a_m^2 \, P_{\rm radar}}
  {k_B \, T_{\rm temp} }
  \right)^{\!\delta/(TB)} 
  \, .
 \label{eq:justRadar}
\end{align}
It is worth noting,
that by employing this estimation entropy in the rate bound,
it is assumed that the estimator
achieves the Cramer-Rao performance.
If the error variance is larger,
then the rate bound is lowered.

\section{Inner Rate Bounds} 

It would be surprising if the performance bound displayed
for the communications multiple-access scenario
in Figure \ref{fig:pentagonOuterBound}
achieved the performance bounds 
of the joint estimation and communications problem.
Here, 
we search for a good
achievable (inner) bounds.
The fundamental system performance
limit lies between these achievable
bounds and the outer bounds found above.
To find these inner bounds,
we hypothesize an idealized receiver
and determine the bounding rates.
To simplify the discussion,
we consider only a single target
with delay $\tau$ 
and gain-propagation-cross-section product $a^2$,
and drop the explicit index to the target.
For example $\sigma^2_{\tau, \rm proc} \rightarrow  \sigma^2_{\rm proc}$.

If $R_{\rm est} \approx 0$
is sufficiently low,
then the communications
operates according to the bound
determined by the isolated communications system,
\begin{align}
  R_{\rm com} 
  &\le B\, 
  \log_2\!\left(1 +  \frac{b^2 \, P_{\rm com}}{\sigma^2_{\rm noise}}\right) 
  \nonumber \\
  &= B\, 
  \log_2\!\left(1 + \frac{b^2 \, P_{\rm com}}{k_B \, T_{\rm temp} \, B}\right) \, .
  \label{eq:comsAlone}
\end{align}

If $R_{\rm com}$
is sufficiently low 
for a given transmit power
then the communications signal can be decoded
and subtracted completely from
the underlying signal,
so that the radar parameters
can be estimated without contamination,
\begin{align}
  R_{\rm com} 
  &\le B\, \log_2\!\left[ 1 + \frac{b^2 \, P_{com}}{\sigma^2_{\rm int+n} }\right] \,
  \label{eq:critCommsRadar}
  \\
  &= B\, \log_2\!\left[ 1 +  \frac{b^2 \, P_{com}}
  { a^2 \, P_{\rm radar} \, \gamma^2 B^2 \, \sigma^2_{\rm proc} 
  + k_B \, T_{\rm temp} \,B  }\right] \, ,
  \nonumber 
\end{align}
where we used Equation \eqref{eq:intPlusNoise}.
In this regime,
the corresponding estimation rate bound $R_{\rm est}$
is given by Equation \eqref{eq:justRadar}.

These two vertices correspond to the points 2 
(associated with Equation \eqref{eq:comsAlone}) and 4 
(associated with Equations \eqref{eq:critCommsRadar} 
and \eqref{eq:justRadar})
in Figure \ref{fig:pentagonOuterBound},
if $R_1$ is interpreted as the estimation rate,
and $R_2$ is interpreted as the communications rate.
An achievable rate lies
within the triangle
constructed by connecting a straight line between these points.

\subsection{Water-filling} 
We hypothesize that we can construct
tighter (larger) inner bounds than 
we constructed in the previous section.
In this section,
we consider a water-filling approach
that splits the total bandwidth into
two sub-bands
and we water fill the communications
power between these bands.
Water filling optimizes
the power and rate allocation
between multiple channels \cite{Cover06,BlissAWC2013}.
For this application,
we separate the band into two frequency channels.
One channel has only communications,
and the other channel is mixed-use
and operates at the SIC rate vertex define by
Equations \eqref{eq:justRadar} and \eqref{eq:critCommsRadar}.

Given some $\alpha$,
that defines the bandwidth separation,
\begin{align}
  B &= B_{\rm com} + B_{\rm mix} \\
  B_{\rm com} &= \alpha \, B 
  \nonumber \\
  B_{\rm mix} &= (1-\alpha) \, B \, ,
  \nonumber 
\end{align}
then we optimize $\beta$
that defines the power utilization,
\begin{align}
  P_{\rm com} &= P_{\rm com,com} + P_{\rm com,mix} \\
  P_{\rm com,com} &= \beta \, P_{\rm com} 
  \nonumber \\
  P_{\rm com,mix} &= (1-\beta) \, P_{\rm com} \, .
  \nonumber 
\end{align}
There are two effective channels
\begin{align}
  \mu_{\rm com} &= \frac{b^2}{k_B \, T_{\rm temp} \, B_{\rm com} }
  \nonumber \\
  &= \frac{b^2}{k_B \, T_{\rm temp} \, \alpha \, B }
  \, ,
\end{align}
for the channel with only communications signal,
and the mixed-use channel
that includes the interference to the communications system
from the radar
\begin{align}
  \mu_{\rm mix} 
   &= \frac{ b^2}{\sigma^2_{\rm int+n} }
   \\
  &= \frac{ b^2}{
  a^2 \, P_{\rm radar} \, (1-\alpha)^2  \gamma^2 B^2 \, \sigma^2_{\rm proc} 
  + k_B \, T_{\rm temp} \, (1-\alpha) \, B  }
  \nonumber
  \, .
\end{align}
The communications power  
is split between the two channels \cite{Cover06,BlissAWC2013},
\begin{align}
  P_{\rm com} &= P_{\rm com,com} + P_{\rm com,mix} 
  \nonumber \\
  &= \left(\alpha \, \nu - \frac{1}{\mu_{\rm com}} \right)^{\!+} 
  + \left( (1-\alpha)\, \nu - \frac{1}{\mu_{\rm mix}}\right)^{\!+} 
  \, .
\end{align}
The critical point (the transition
between using one or both channels
for communications)
occurs when
\begin{align}
  (1-\alpha)\, \nu - \frac{1}{\mu_{\rm mix}} &= 0
  \nonumber \\
  P_{\rm com} &= \alpha \, \nu - \frac{1}{\mu_{\rm com}} \, ,
\end{align}
so
both channels are used if
\begin{align}
  P_{\rm com}& \ge \frac{\alpha}{(1-\alpha) \, \mu_{\rm mix}}- \frac{1}{\mu_{\rm com}} 
  \, .
  \label{eq:dualUse}
\end{align}

If the communications-only channel is used exclusively
for communications,
then $P_{\rm com} = P_{\rm com,com}$.
If both channels are employed for communications then
\begin{align}
  P_{\rm com,com} 
  &= \alpha \, \nu - \frac{1}{\mu_{\rm com}}
  \nonumber \\
  P_{\rm com,mix}&= (1-\alpha ) \, \nu - \frac{1}{\mu_{\rm mix}} \, ,
\end{align}
and thus when Equation \eqref{eq:dualUse}
is satisfied
\begin{align}
  P_{\rm com} &=  \alpha \, \nu - \frac{1}{\mu_{\rm com}} 
  +(1-\alpha ) \,  \nu - \frac{1}{\mu_{\rm mix}}
  \nonumber \\
  \nu &= 
  \left(P_{\rm com} + \frac{1}{\mu_{\rm com}} + \frac{1}{\mu_{\rm mix}}\right) 
  \, .
\end{align}
The value of power fraction $\beta$
is then given by
\begin{align}
  \beta &= \frac{P_{\rm com,com}}{P_{\rm com}} 
  \nonumber \\
  &=\frac{\alpha \, \nu - \frac{1}{\mu_{\rm com}}}{P_{\rm com}} 
  \nonumber \\
  &=\frac{\alpha \, 
  \left(P_{\rm com} + \frac{1}{\mu_{\rm com}} + \frac{1}{\mu_{\rm mix}}\right) 
  - \frac{1}{\mu_{\rm com}}}{P_{\rm com}}   
  \nonumber \\
  &=\alpha + \frac{1}{P_{\rm com}}\left(
  \frac{\alpha-1}{\mu_{\rm com}} + \frac{\alpha}{\mu_{\rm mix}}\right)  \, ; 
  \nonumber \\
  &\quad \mbox{ when } \, 
  P_{\rm com}\ge \frac{\alpha}{(1-\alpha) \, \mu_{\rm mix}}- \frac{1}{\mu_{\rm com}} 
  \, .
\end{align}
The resulting communications rate bound
in the communications-only subband
is given by
\begin{align}
  R_{\rm com,com} 
  &\le
  B_{\rm com} 
  \log_2\left[ 1 + \frac{P_{\rm com, com} \, b^2}{k_B \, T_{\rm temp} \, B_{\rm com} }\right] 
  \nonumber \\
  &\le 
  \alpha \, B
  \log_2\left[ 1 + \frac{\beta \, P_{\rm com} \, b^2}{k_B \, T_{\rm temp} \, \alpha \, B }\right] 
  \, .
\end{align}
If $P_{\rm com} < 1/\mu_{com} - 1/\mu_{mix}$
then $R_{com,mix} = 0$
because no communications power
is allocated to the ``mixed" use channel,
otherwise the mixed use communications rate inner bound
is given by
\begin{align}
  R_{\rm com,mix} 
  &\le
  B_{\rm mix} 
  \log_2\left[ 1 + 
  \frac{b^2 \, P_{\rm mix} }
  {\sigma^2_{\rm int+n}}
  \right] 
  \nonumber \\
  &=
  (1-\alpha) \, B \, 
  \log_2\left[ 1 + 
  \frac{b^2 \, (1-\beta) \, P_{\rm com}}
  {\sigma^2_{\rm int+n}}
  \right] 
  \\
  \sigma^2_{\rm int+n}
  &= a^2 \, P_{\rm radar} \, (1-\alpha)^2 \, \gamma^2 \, B^2 \, \sigma^2_{\rm proc} 
  +k_B \, T_{\rm temp} \, (1-\alpha) \, B 
  \nonumber 
\end{align}
The corresponding radar estimation rate
inner bound
is then given by
\begin{align}
  R_{\rm est} 
  &\le 
  B_{\rm mix} \log_2\!\left( 1 + 
  \frac{ \sigma_{proc}^2 \, \gamma^2 \, B_{\rm mix} \, (TB_{\rm mix}) \,
   a^2 \, P_{\rm radar}}{k_B \, T_{\rm temp} }
  \right)^{\!\delta/(TB_{\rm mix})} 
  \nonumber \\
  &= 
  (1-\alpha)\,B \log_2\!\left( 1 + 
  {\rm SNR_{radar}}
  \right)^{\!\delta/([1-\alpha]\,TB)} 
\end{align}
\begin{align}
 {\rm SNR_{radar}} &= 
  \frac{ \sigma_{proc}^2 \, \gamma^2 \, (1-\alpha)\,B \, ([1-\alpha]\,TB) \,
   a^2 \, P_{\rm radar}}{k_B \, T_{\rm temp} }
  \, .
\end{align}
We assume that $[1-\alpha]\,TB$,
which is the waveform integration,
 is held constant
as $\alpha$ is varied
so $R_{\rm est}$ is given by
\begin{align}
  R_{\rm est} 
  &\le
  (1-\alpha)\,B \log_2\!\left( 1 + 
  \frac{ \sigma_{proc}^2 \, \gamma^2 \, (1-\alpha)\,B \, \kappa \,
   a^2 \, P_{\rm radar}}{k_B \, T_{\rm temp} }
  \right)^{\!\delta/\kappa} 
  \, ,
\end{align}
where waveform integration is denoted $\kappa = (1-\alpha)\,TB$.
For some very large value of $\alpha$,
corresponding to a very small radar subband,
the problem is no longer self consistent because $T>T_{\rm rpi}$.

\subsection{Examples} 
\begin{figure}[htb]
\centering
   \includegraphics[width=3.1in]{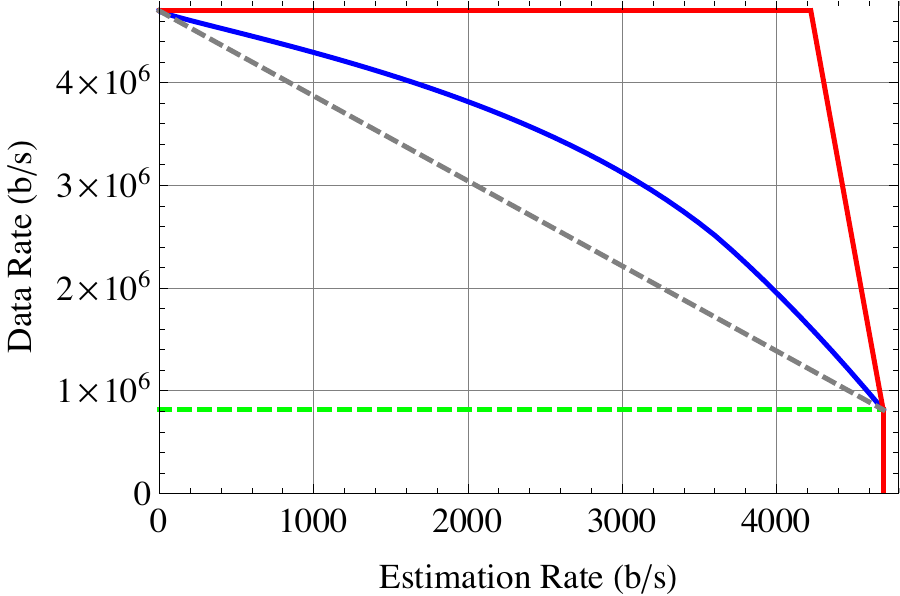}
   \caption{
   Data rate and estimation rate bounds.
   Outer bounds on communications and radar
   are indicated by the red lines.
   Successive interference cancellation (SIC) bound for the communications rate
   is indicated by the green dashed line.
   The linear interpolation between SIC vertex and the 
   radar-free data rate bound
   is indicated by the gray dashed line.
   The water-filling inner bound is indicated by the blue line.
  \label{fig:estRateCommRate}
  }
\end{figure}

In Figure \ref{fig:estRateCommRate},
we display an example
of inner bounds on performance.
The parameters
used in the example are displayed
in Table \ref{tab:exampleParameters}.
It is assumed that the communications
system is received through an antenna sidelobe,
so that the radar and communications
receive gain are not identical.
In the figure,
we indicate a outer bound in red.
We indicate in green,
the bound on successive interference cancellation (SIC),
presented in Equation \eqref{eq:critCommsRadar}.
The best case system performance
given SIC
is at the vertex 
(at the intersection of the green and red lines),
which is determined 
by the joint solution of Equations  \eqref{eq:critCommsRadar} 
and \eqref{eq:justRadar}.
The inner bound that linearly interpolates
between this vertex and the radar-free communications
bound in Equation \eqref{eq:comsAlone}
is indicated by the gray dashed line.
The water-filling bound is indicated by the blue line.
The water-filling bound is not guaranteed to be convex.
The water-filling bound is not guaranteed to be greater
than the linearly interpolated bound.
In general,
the inner bound is produced
by the convex hull of all contributing inner bounds.
In the example,
we see that the water-filling bound
exceeds the linearly interpolated bound.

\begin{table}[htdp]
\caption{Parameters for Example Performance Bound.}
\begin{center}
\begin{tabular}{|c|c|}
\hline
Parameter & Value \\
\hline \hline 
Bandwidth & 5~MHz \\
Center Frequency & 3~GHz \\
Temperature & 1000~K \\
Communications Range & 10~km \\
Communications Power & 20~dBm \\
Communications Antenna Gain & 0~dBi \\
Radar Target Range & 100~km \\
Radar Antenna Gain & 30~dBi \\
Radar Power & 1~kW \\
Target Cross Section & 10~m$^2$ \\
Target Process Standard Deviation & 100~m \\
Time-Bandwidth Product & 100 \\
Radar duty factor & 0.01 \\
\hline
\end{tabular}
\end{center}
\label{tab:exampleParameters}
\vspace{-5mm}
\end{table}
%

\section{Conclusion} 

In this paper,
we provide a novel approach
for producing joint radar and communications
performance bounds.
An achievable inner bound
based on a water-filling approach
is developed and an example is presented.
This is an initial
investigation.
There are a range of potentially significant improvements 
to the inner bounds,
and potentially interesting extensions
to the scenarios to which these bounds may be applied.


\bibliographystyle{IEEEtran}
\bibliography{../blissPaper}

\end{document}